\documentclass[traditabstract]{aa}
\usepackage{txfonts}
\usepackage[english]{babel}
\usepackage{t1enc}
\usepackage[ansinew]{inputenc}
\usepackage{graphicx}
\usepackage{todonotes}
\usepackage{subfig}
\usepackage{lscape}
\usepackage{natbib}
\usepackage{footmisc}
\usepackage{url}
\usepackage{dingbat}

\begin{document}

\title{A multi-wavelength view of the Galactic center dust ridge reveals little star formation}

\author{K. Immer\inst{1,2} \and K.~M. Menten\inst{1} \and F. Schuller\inst{3,1} \and D.~C. Lis\inst{4}}

\institute{Max-Planck-Institut f\"ur Radioastronomie, Auf dem H\"ugel 69, D-53121 Bonn, Germany
\and
Harvard-Smithsonian Center for Astrophysics, 60 Garden Street, 02138 Cambridge, MA, USA
\and
ESO, Alonso de Cordova 3107, Casilla 19001, Santiago 19, Chile
\and
California Institute of Technology, Cahill Center for Astronomy and Astrophysics 301-17, Pasadena, CA 91125, USA
}

\date{Received / 
      Accepted }

\abstract{The Galactic center dust ridge consists of a narrow string of massive condensations identified in submillimeter dust continuum emission. To determine whether new high-mass stars are forming in this region, we performed new observations at 870 $\mu$m with the Atacama Pathfinder Experiment telescope and at 8.4 GHz with the Very Large Array. We complement our data with recent maser and mid-infrared results. The ridge's clouds are dark at mid-infrared wavelengths, indicating the presence of cold, high column density material. In combination with existing temperature measurements in the dust ridge, we determined the masses of the largest clouds. The results show that the dust ridge contains a very massive reservoir of molecular material. We find five radio sources at 8.4 GHz in the general dust ridge vicinity but outside of the dust ridge clouds, which are probably all excited by massive young stars, whose properties we constrain. Our observations exclude the existence of zero age main sequence stars with spectral types earlier than B0.5 within the dust ridge clouds. The only indication of ongoing high-mass star formation inside the clouds are class II methanol masers that are found in two of the clouds. Except for a weak water maser, found in previous observations, no signs of star formation are detected in the massive cloud M0.25+0.012.}

\keywords{Galaxy: center -- Stars: formation -- \ion{H}{II} regions}

\authorrunning{K. Immer et al.}
\titlerunning{The Dust Ridge}

\maketitle

\section{Introduction}
\label{Intro}

The dust ridge \citep[named by][]{Lis1994a} is an accumulation of clumpy dust condensations near the Galactic Center, which are located along an arc-like narrow ridge, connecting the radio continuum sources G0.18$-$0.04 and Sgr B1 at Galactic longitude $l \approx$ 0\fdg{5}. 

To determine the dust temperature of the molecular clouds, \citet{Lis1998} and \citet{Lis1999} imaged the region using the Long Wavelength Spectrometer (LWS) aboard the Infrared Space Observatory (ISO) at wavelengths from 45 to 175 $\mu$m. They found that the distribution of the emission at wavelengths longer than 70 $\mu$m was well correlated with the submillimeter continuum emission whereas the clouds were seen in absorption against the general Galactic Center background at shorter wavelengths. This implies that the clouds are much colder than and located in front of the warm dust responsible for the emission at 70 $\mu$m and that they must have a high continuum opacity at 70 $\mu$m. In addition, \citet{Lis1999} estimated the temperature at six points in the clouds by fitting the observed spectral energy distributions (SEDs) with a two-component grey-body model. They found out that the bulk of the dust has temperatures between 13 and 20 K but that also a small amount of dust with a higher temperature, which potentially corresponds to emission at the surface of the clouds, is necessary to accurately model the SEDs. Ammonia observations of M0.25+0.012 by \citet{Guesten1981} suggest that the gaseous material has higher temperatures than the dust, probably being heated by cosmic rays.

\citet{Molinari2011} interpret the dust ridge as a quarter of a 100 pc size elliptical twisted ring of molecular material around the Galactic Center. Their Herschel observations of the central molecular zone show the dust ridge in absorption at 70 $\mu$m and in emission at 250 $\mu$m, confirming the low temperatures of the dust ridge clouds, estimated by \citet{Lis1999}. The column density map of atomic hydrogen, derived from the 250 $\mu$m map,  shows strong peaks at the position of the dust ridge clouds, indicating high column densities along the line of sight to the clouds.

\citet{Lis1998} determined the distribution of the CO (2--1) and the HCO$^+$ (3--2) emission at the location of M0.25+0.012, the largest cloud in the dust ridge. The line widths of these molecules are very broad ($\sim$ 30 kms$^{-1}$ FWHM), similar to the observed line widths in giant molecular clouds (GMCs) in the Galactic Center, indicating that M0.25+0.012 is probably located at the same distance as these \citep[i.e., at 8.5 kpc, the Galactic Center distance;][]{Ghez2008}. 

Furthermore, Lis et al. detected a large velocity gradient across the southern part of the cloud indicating streaming gas motions or several spatially overlapping velocity components. This could be a sign of an occurring cloud collision. As the cloud has not yet been warmed up by shocks, it is likely that the collision is at an early stage, which could trigger substantial high mass star formation inside the cloud in the future.

M0.25+0.012 was recently studied by \citet{Longmore2012}, who present data from near-infrared to millimeter wavelengths. They determined global properties such as dust mass, dust temperature and column density for this cloud and estimated the distance to the cloud from extinction measurements. Their results are consisted with a Galactic Center distance of 8.5 kpc and the temperature measurements of \citet{Lis1999}. They conclude that this cloud could be a precursor for an Arches-like young massive cluster.

In this paper, we present new observations of the dust ridge at 870 $\mu$m with the Atacama Pathfinder Experiment (APEX) telescope and at 8.4 GHz with the Very Large Array (VLA). In Section \ref{Obs}, we describe the observations and the data reduction of the two data sets. Section \ref{SFDR} gives a comprehensive picture of the dust ridge at infrared, submillimeter and radio wavelengths. In the last section, we summarize our results.

\section{Observation and Data Reduction}
\label{Obs}

\setlength{\tabcolsep}{0.9mm}

\begin{table}
	\centering
		\begin{tabular}{|l|c|l|l|}\hline
    Target name 	& Total on-source  	& \multicolumn{2}{|c|}{Pointing center}\\
                		&  time 			& R.A. (J2000)	& Dec (J2000)\\
                		&     [s]  		         & [hh mm ss.ss] 	     	& [dd $\arcmin\arcmin$ $\arcsec\arcsec.\arcsec\arcsec$]\\ \hline
    B1328+307    	& 200           		& 13 31 08.29  		& +30 30 33.04\\
    B1730$-$130 & 1820          		& 17 33 02.69		& $-$13 04 49.55\\
    POS1        	& 3720          		& 17 46 08.59  		& $-$28 42 37.49\\
    POS2        	& 4180          		& 17 46 10.17 		& $-$28 38 07.37\\
    POS3		& 4640          		& 17 46 22.09 		& $-$28 34 36.50\\\hline
		\end{tabular}
		\caption{Integration times (Col. 2) and pointing centers (Col. 3 and 4) of the five fields (Col. 1) observed with the VLA at 8.4 GHz.}
		\label{PoiCenter+TOS8.4GHz}
\end{table}

Three target positions (POS1, POS2, POS3) were observed in 1993 June at 8.4 GHz in the BC hybrid configuration with the NRAO Very Large Array (VLA) covering almost the whole dust ridge. Data were recorded in two intermediate frequency (IF) bands with opposite (right and left) circular polarization, RCP and LCP.  Each band was 50 MHz wide (46 MHz effective bandwidth). They were centered at 8.4149 GHz (IF 1 = RCP) and 8.4649 GHz (IF 2 = LCP).

Fourteen scans on B1730$-$130, the phase calibrator, were recorded before, as well as after the observations of the target sources, resulting in a total on-source time of half an hour. B1328+30, the amplitude calibrator, was observed only once for 200 s. The total integration time of each target source, comprising 12 or 13 individual scans, is larger than one hour. The accurate values are listed together with the pointing centers of the observed fields in Table \ref{PoiCenter+TOS8.4GHz} (Col. 2, 3, and 4; Col. 1 gives the target names).

After ``flagging'' time ranges in which certain baselines contained obviously bad data and the amplitude and phase calibration, several self-calibration cycles were repeated and the final images, restored with an elliptical Gaussian beam of $\sim3\farcs{6}$~$\times$~$\sim2\farcs{5}$ FWHM, were then corrected for primary beam attenuation.  At our observing frequency, the primary beam has a FWHM size of $5\farcm{4}$, which determines the field of view. 

The images resulting from the POS1, POS2, and POS3 pointings have rms values of ~0.25, ~0.15, and ~0.25 mJy~beam$^{-1}$. We will compare our radio observations with a VLA map, taken at 1.4 GHz by \citet{Yusef-Zadeh2004} with a 1-$\sigma$-sensitivity of $\sim$12 mJy~beam$^{-1}$ (in the dust ridge region) in a $\sim$30$\arcsec$ beam, and a VLA catalog of discrete sources at 5 and 1.4 GHz from \citet{Becker1994}. Sources in this catalog are detected down to 9 mJy at 5 GHz and 10 mJy at 1.4 GHz with a resolution of 4--5$\arcsec$. Our VLA observations have higher resolution and are a factor of 10--40 more sensitive than these comparable observations.

The submillimeter images of the dust ridge at 870 $\mu$m were taken in the framework of the ATLASGAL survey \citep{Schuller2009} with the APEX\footnote{The APEX project is a collaborative effort between the Max Planck Institute for Radioastronomy, the Onsala Space Observatory, and the European Southern Observatory.} telescope \citep{Guesten2006}. The resolution of the submillimeter maps is $19\farcs{2}$, and the map has been projected on 2$\arcsec$ pixels. The rms between pixels is in the range 40--60 mJy beam$^{-1}$ in regions with no emission. As compared to the data presented in \citet{Schuller2009}, more coverages of the full region have been combined to produce the map that we use in the present paper. 
The field was covered by combining eight on-the-fly maps, each roughly orthogonal to the Galactic plane. Each map was executed five times with slightly changing orientation to minimize stripping effects. Also, we used a slightly different reduction pipeline specially developed in order to recover as much extended emission as possible. The main idea is to build the map in an iterative way, starting with the assumption that there is no emission on the edges of the map (at $b = \pm1$). Then, the processing is similar to the one described in Schuller et al. (2009), but a major difference concerns the correlated noise removal. As in the original pipeline, a median value of all bolometer signals is computed at each integration, and is considered as sky signal, $S$, to be subtracted from all signals. But, in addition, the bolometer where the signal is equal to the median value $S$ is identified. The position observed by this bolometer at this time is then uniquely determined. The procedure then looks for the intensity at this position in the map used as input model (i.e. the map resulting from the previous iteration). If there is significant emission, $E$, then $S$ is not subtracted from all signals, but only the difference $S$-$E$ is considered as sky, and is subtracted. This way, real emission is preserved from one iteration to the next.

%For a quantitative estimate of the loss in peak and integrated flux density as a function of %source size, we refer to the simulations presented by \citet{Belloche2011}, who performed a %similar data processing on LABOCA data.

We also made use of Spitzer/IRAC images at 3.6, 4.5, 5.8, and 8.0 $\mu$m that were obtained as part of the Galactic Legacy Infrared Mid-Plane Survey Extraordinaire (GLIMPSE) \citep{Churchwell2009}.
%as well as the Spitzer/MIPS image at 24 $\mu$m from the MIPSGAL Galactic Plane survey

\section{High-Mass Star Formation in the Dust Ridge}

\label{SFDR}

\subsection{Dust Emission in the Dust Ridge}
\label{DustDR}

\begin{figure*}
	\centering
   \includegraphics[angle=270,width=18cm]{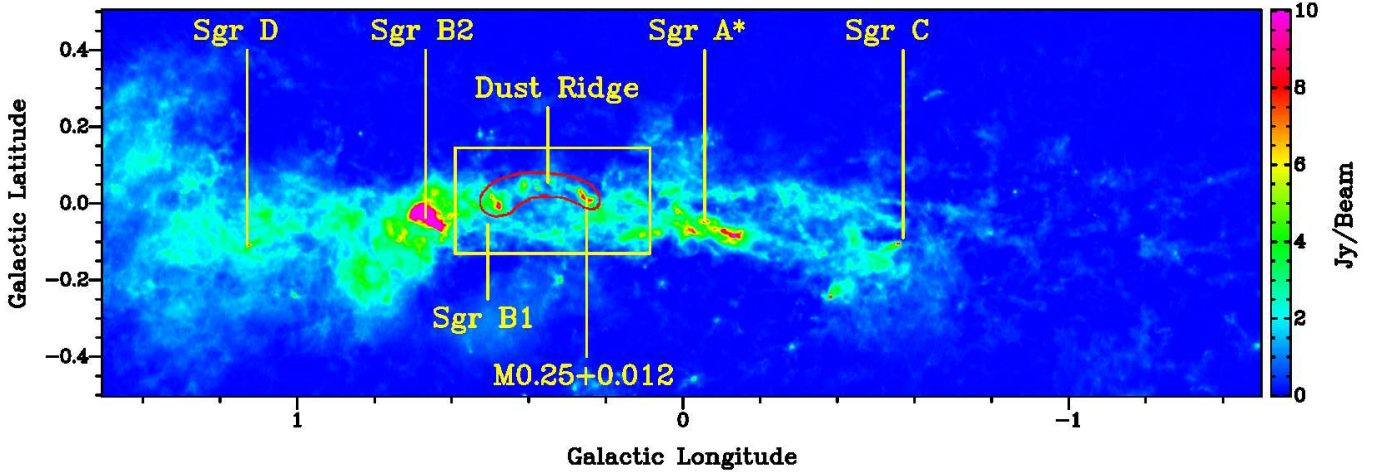}
		\caption{870 $\mu$m dust continuum emission of the central molecular zone. The figure shows the position of the dust ridge (red contour) and the large molecular cloud M0.25+0.012 in the central molecular zone in relation to other famous molecular complexes: Sgr A*, Sgr B1, Sgr B2, Sgr C, and Sgr D. The region within the yellow box is enlarged in Fig. \ref{DRSubmmTM}.}
	\label{CMZSubmm}
\end{figure*}

\begin{figure*}
    \centering
    \includegraphics[width=18cm]{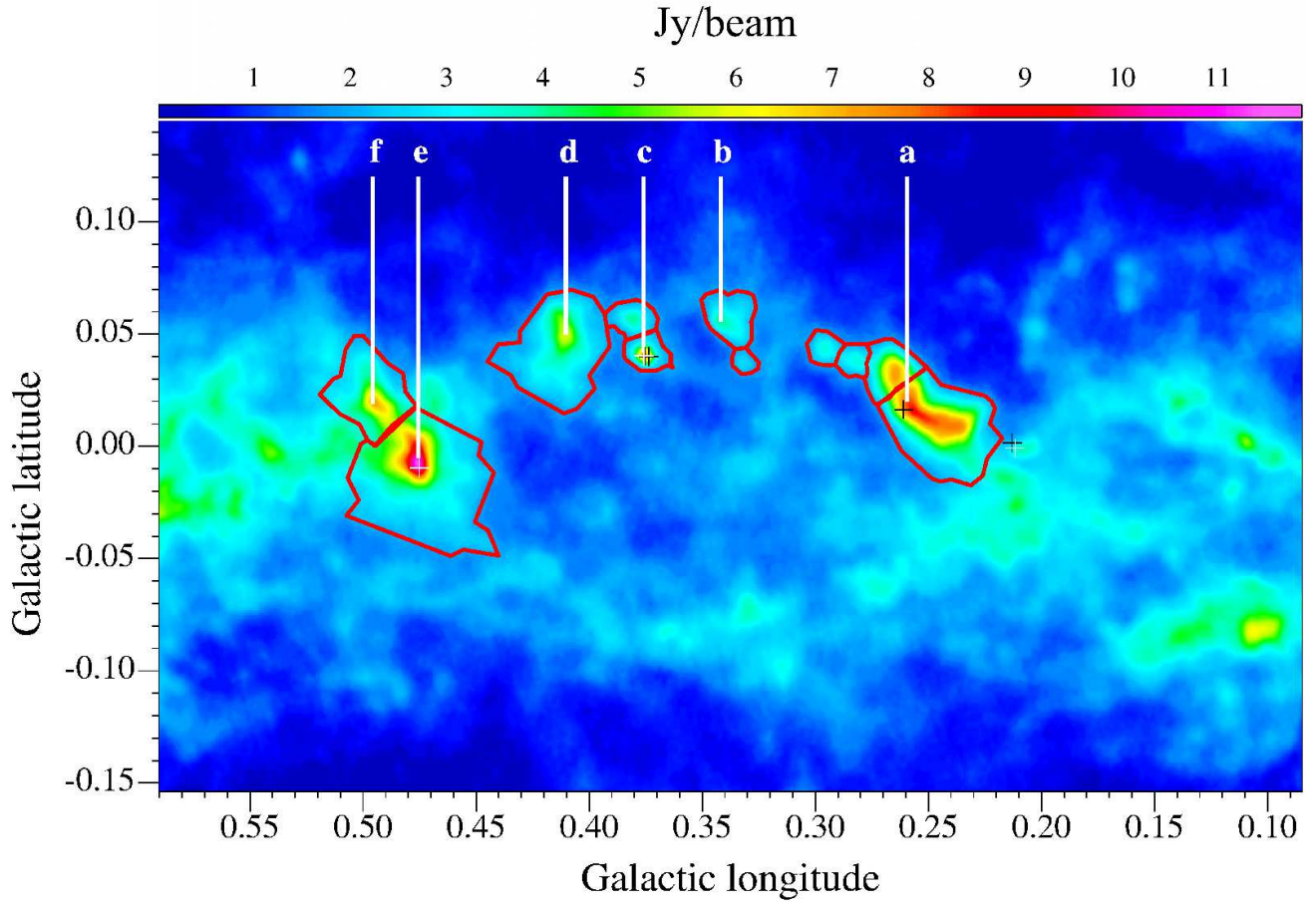}
		\caption{Dust continuum emission of the dust ridge (arc of dust condensations bent to the north) at 870 $\mu$m. The contours show the sizes of the molecular clouds over which the emission was integrated for the determination of the cloud masses. Water and methanol masers are marked with black and white crosses, respectively. The letters mark the positions where \citet{Lis1999} determined the temperatures of the dust.}
	\label{DRSubmmTM}
\end{figure*}

Fig. \ref{CMZSubmm} presents an image of the central molecular zone in 870 $\mu$m emission around the Galactic center  (|l|~<~1$\fdg{5}$, |b|~<~0$\fdg{5}$). The dust ridge is indicated, as well as the famous molecular complexes Sgr A*, Sgr B1, Sgr B2, Sgr C, and Sgr D.
From multi-wavelength submillimeter observations one concludes that most of this emission is produced by dust associated with these complexes.

Despite the better sensitivity of the Herschel Hi-GAL \citep{Molinari2010} project, a comparison of our Fig. \ref{CMZSubmm} with Herschel observations at 250 $\mu$m \citep[see Fig. 2 in][]{Molinari2011} shows a very similar distribution of the dust emission. The Hi-GAL observations cover a wavelength range from 70 to 500 $\mu$m, showing dust of higher temperature than our ATLASGAL observations. Since the Hi-GAL data and the ATLASGAL observations have comparable resolutions, they can be used in combination to derive spectral energy distributions, and thus the temperatures, of the molecular clouds.

\begin{table}
 \centering
  \begin{tabular}{|c|c|c|c|c|}\hline
	Position 	& T$_{Dust}$ [K] 	& F$_{870 \mu m, Int.}$ [Jy]	& M$_{Cloud}$ [10$^3$ M$_\odot$] & D$_{Cloud}$ [$\arcsec$]\\ \hline
         a (M0.25+0.012) 	& 18 			& 310		& 141 & 207\\
         b 				& 22 			& 39			& 13 & 96\\
         c 				& 20 			& 46 			& 18 & 96\\
         d 				& 17 			& 145 		& 72 & 171\\
         e 				& 17 			& 307 		& 153 & 220\\
         f 				& 15 			& 118  		& 72 & 133\\\hline
   \end{tabular}
   \caption{Sub-condensations in the dust ridge in the nomenclature of \citet{Lis1999} (Col. 1; see also Fig. \ref{DRSubmmTM}). Listed are derived dust temperatures (Col. 2), integrated flux densities of the 870 $\mu$m emission (Col. 3), and total masses (Col. 4). The equivalent diameter D (Col. 5) is an estimate of the cloud size, derived from the area of the clumps given by the ClumpFind algorithm.}
   \label{Masses-Dustclouds}
\end{table}

Fig. \ref{DRSubmmTM} shows the emission of the dust ridge at 870 $\mu$m (red contour in Fig. \ref{DRSubmmTM}). The ridge consists of several dust condensations and extends, from Sagittarius B1 to M0.25+0.012, over 0.25$\degr$ \citep[40 pc at the Galactic center distance of 8.5~kpc;][]{Ghez2008}. Several weaker dust condensations are detected in the south of the dust ridge which form a projected ellipse with the dust ridge clouds.
The letters indicate the condensations identified by \citet{Lis1999}, for which they determined dust temperatures from ISO data. The lower longitude end of the ridge is located at a projected distance of around 30 pc from the Galactic Center.
%, assuming a distance %to the Galactic Center of 8.5 kpc \citep{Ghez2008}.  
Fig. \ref{DRSubmmTM} shows that the 870 $\mu$m appearance of the dust ridge region is dominated by diffuse emission with a mean value of $\sim$1.7 Jy~beam$^{-1}$. To separate the diffuse emission from the emission of core regions and identify clumps of emission, we used the CUPID package of the starlink software with the established ClumpFind algorithm \citep{Williams1994}. We chose 1.7 Jy~beam$^{-1}$ as the lowest contour level and 0.3 Jy~beam$^{-1}$ (5$\cdot$rms) as the gap between contour levels. The impact of choosing different ClumpFind levels is discussed later in the text. The red contours in Fig. \ref{DRSubmmTM} show the boundaries of the identified clumps. Clouds a, b, and c are subdivided in several clumps. To quantify the cloud sizes, we assigned equivalent diameters based on the clump areas given by the ClumpFind algorithm.

To determine if the dust ridge clouds are gravitationally bound, the total masses of the clouds from molecular gas and dust and their virial masses have to be compared. The virial mass of a cloud is proportional to the radius of the cloud and the linewidth of the molecular gas squared. However, our submillimeter observations do not provide velocity information of the cloud gas, prohibiting the determination of the virial mass. In a following paper, we study the kinematics of the cloud gas using data from the Millimetre Astronomy Legacy Team 90 GHz Survey \citep[MALT90,][]{Foster2011} and recent APEX spectral line observations to investigate whether the clouds are gravitationally bound. \citet{Longmore2012} showed that cloud a (M0.25+0.012) is indeed gravitationally bound. For this paper, we can only assume that the other outlined clouds are also gravitationally bound and not only line-of-sight superpositions. 
Assuming further that the dust temperature T$_{D}$ does not change significantly within each cloud, we can determine the projected masses of the clouds M$_{Cloud}$ with the formula \citep[following][]{Hildebrand1983}
\[
M_{Cloud} = \frac{d^{2}F_{870 \mu m, Int.}R}{B_{870 \mu m}(T_{D})\kappa_{870 \mu m}}
\]

The integrated flux densities F$_{870 \mu m, Int.}$ of the six clouds are measured by adding up all pixel flux density values within the boundaries of each clump. For the clouds a, b, and c, we summed the integrated flux densities of all clumps belonging to the same cloud. For the gas-to-dust ratio, $R$, the opacity at 870 $\mu$m $\kappa_{870 \mu m}$, and the distance to the clouds, $d$, values of 100, 0.185 m$^2$ kg$^{-1}$, and 8.5 kpc, respectively, were assumed. The results are shown in Table \ref{Masses-Dustclouds} with Column 2 and 3 giving the dust temperatures and the integrated flux density values of the different clouds. The last column contains the determined total masses of the clouds. The mass of cloud a (M0.25+0.012) is consistent with the results of \citet{Longmore2012}. 

The largest uncertainty in the mass determination arises from the determination of the cloud boundaries. Depending on the lowest contour level of the ClumpFind algorithm, the sizes of the clouds change slightly, thus, including more or less diffuse emission at the boundaries of the clouds. For example, a change of the lowest contour level from 1.7 Jy~beam$^{-1}$ to 1.8 Jy~beam$^{-1}$ decreases the masses of the larger clouds by 1--5 \% and the smaller clouds by $\sim$10\%. The equivalent diameter decreases by 1--8 \% and we can conclude that the size of the clouds is robust against small changes in the input level. 

As mentioned in Section \ref{Obs}, the data reduction process was developed for a better extended emission recovery. However, we cannot exclude that some uniform extended emission is still filtered out by the data reduction process. The masses we derive are therefore only lower limits of the real masses. Furthermore, the gas-to-dust-ratio in the central molecular zone may be lower than the standard value of 100 (determined at solar metallicity) due to the negative metallicity gradient in the disk \citep{Balser2011}. Since no metallicity measurements exist for the dust ridge region, we adopted the commonly used value of 100 for the gas-to-dust ratio, but we note that the total masses of the clouds may be lower. We estimate the uncertainty of the total cloud masses to be of the order of $\sim$20\%.

The total mass of the dust ridge adds up to $\sim$ 5 $\cdot$ 10$^5$ M$_{\sun}$, which makes this region one of the most massive dust and gas reservoirs in the vicinity of the Galactic center. Therefore, probably an important part of the future star formation in the central molecular zone will take place in this region.

\subsection{Maser Sources in the Dust Ridge}
\label{MaserDR}

\citet{Lis1994b} conducted an observation of the cloud M0.25+0.012 at 22.2 GHz with the VLA in the B/C hybrid configuration. They found a weak water maser near the peak of the submillimeter emission with an isotropic 22 GHz line luminosity of $1.5 \cdot 10^{-6}$ L$_\odot$. They deduced that this maser is most likely associated with a deeply embedded intermediate or low-mass young stellar object. 

A recently conducted search for class II methanol masers at 6.7 GHz \citep{Caswell2010} showed the existence of three methanol masers in the dust ridge region. They are located in clouds c and e, as well as in the south-west of M0.25+0.012. At the positions of the first two methanol masers also water masers have been detected \citep{Forster1999, Valdettaro2001}. The positions of the water and methanol masers are indicated with black and white crosses, respectively, in Fig. \ref{DRSubmmTM}.

The existence of the methanol masers shows that high-mass star formation is taking place in parts of the dust ridge. However, besides the weak water maser, there were no other masers found in the most massive cloud M0.25+0.012.\\ \

\subsection{Radio Sources in the Dust Ridge}

\begin{figure*}[htbp]
     \centering
      \subfloat[Dust Ridge]{\includegraphics[angle=270,width=18cm]{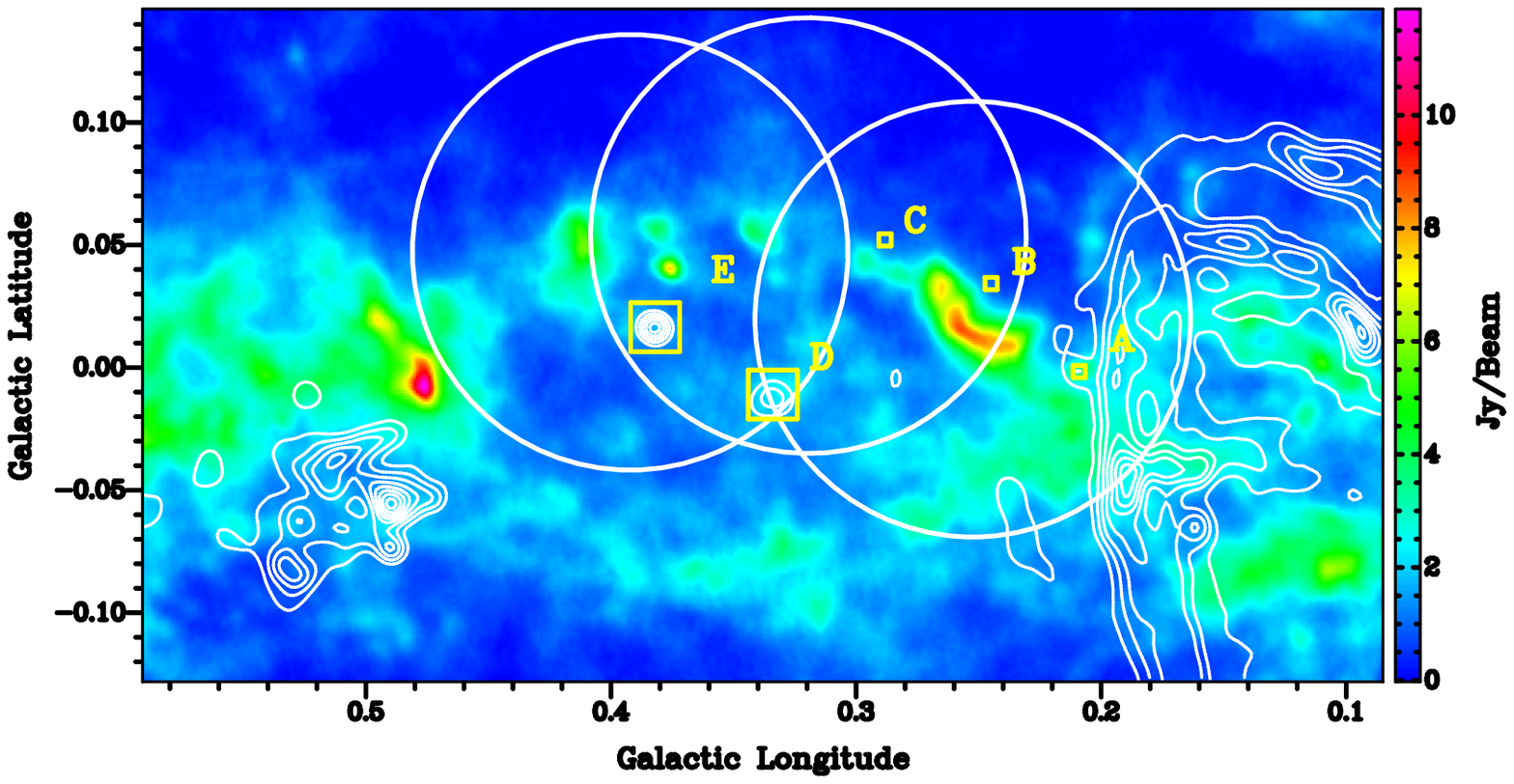}}\\
     \subfloat[Source A]{\includegraphics[angle=270,width=6.3cm]{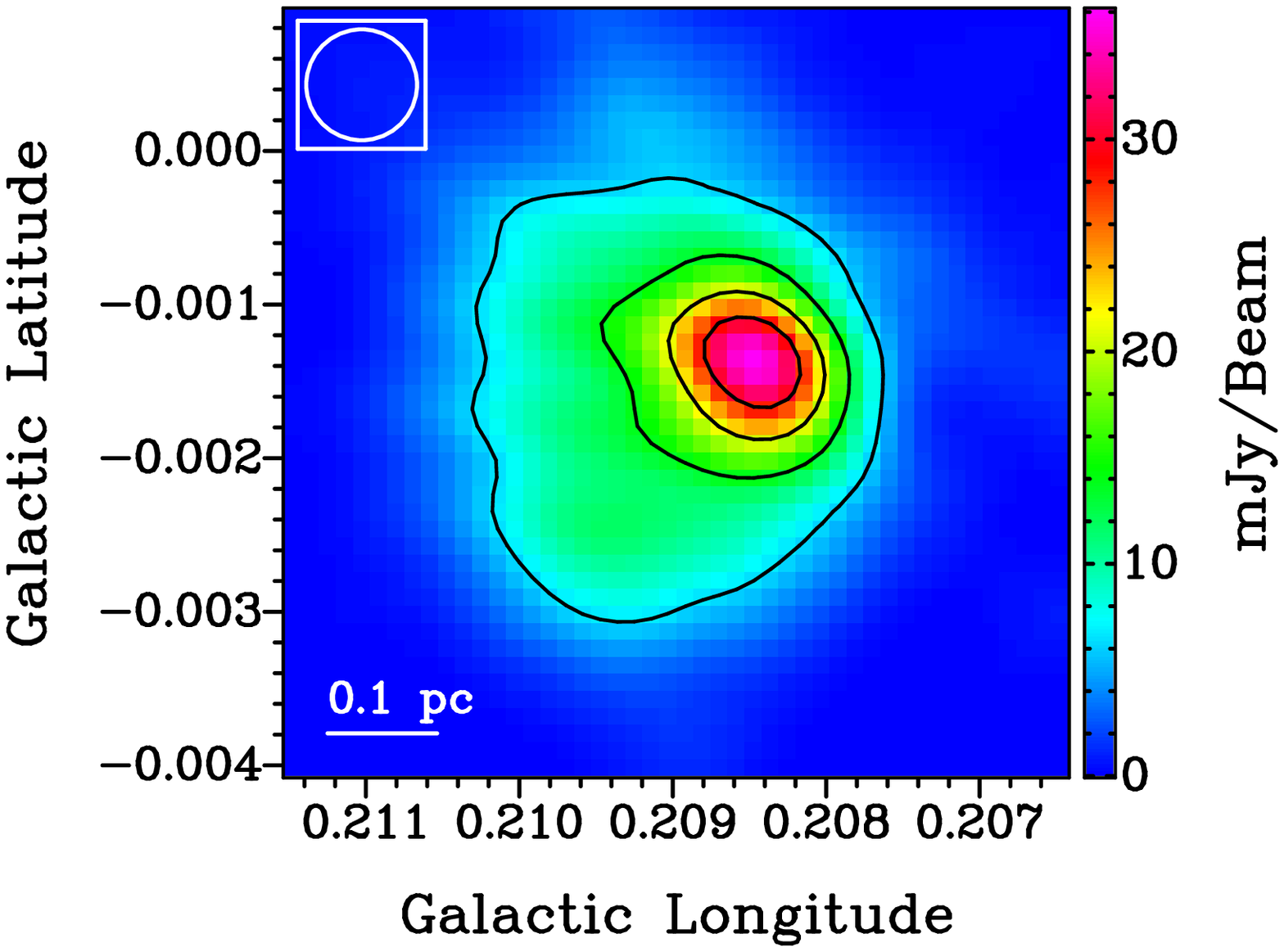}}
     \subfloat[Source B]{\includegraphics[angle=270,width=6cm]{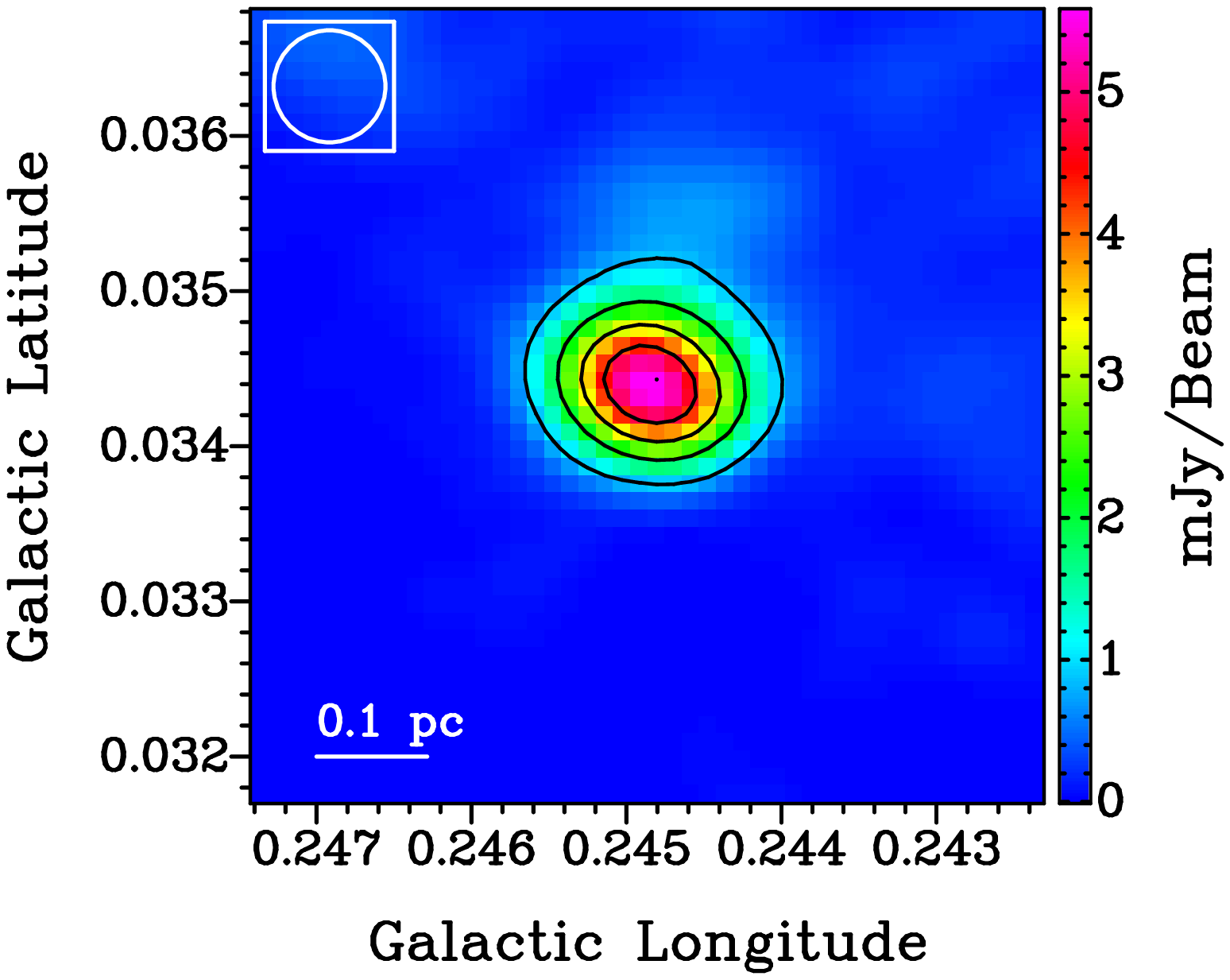}}	
     \subfloat[Source C]{\includegraphics[angle=270,width=6cm]{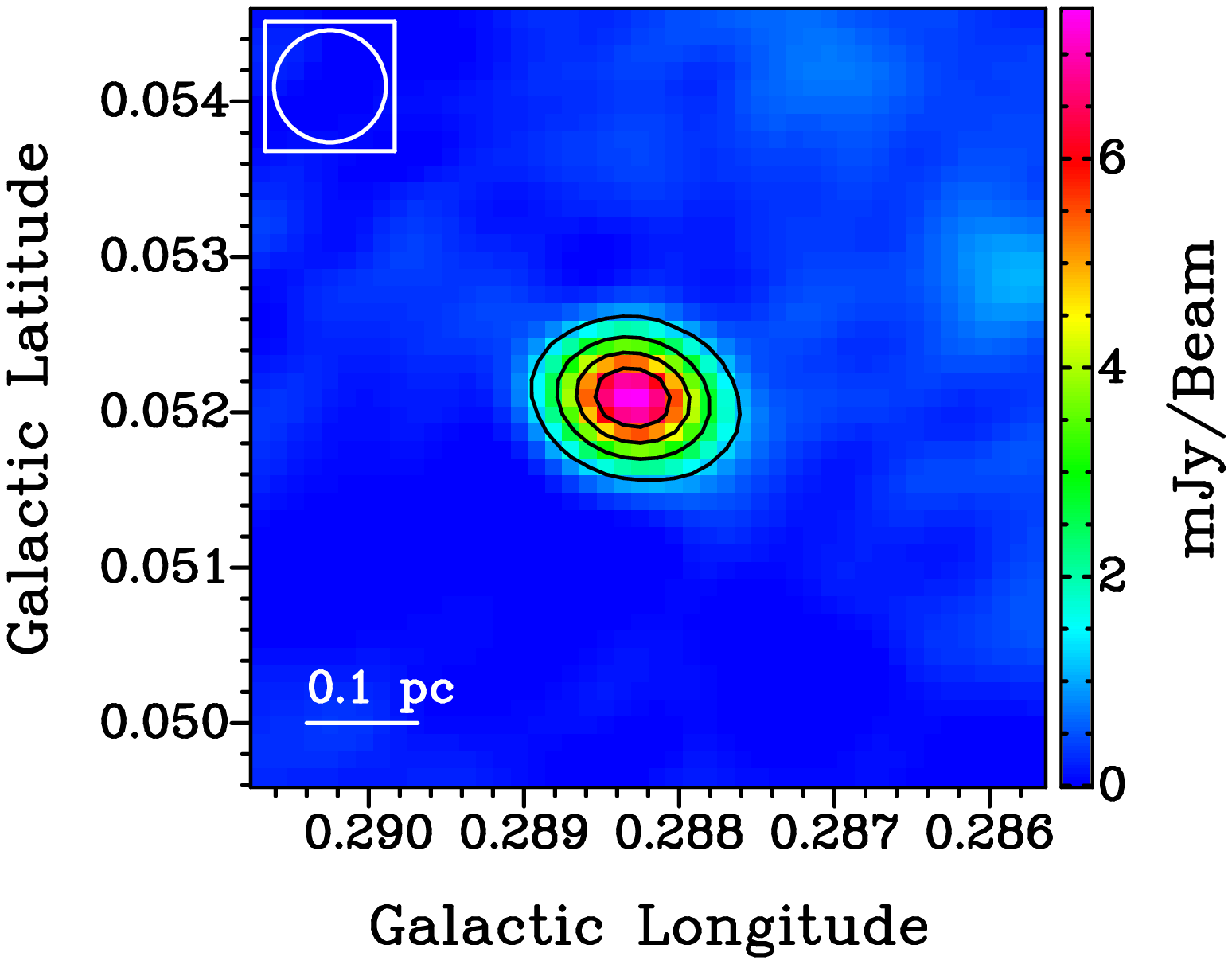}}\\	   	
     \subfloat[Source D]{\includegraphics[angle=270,width=6cm]{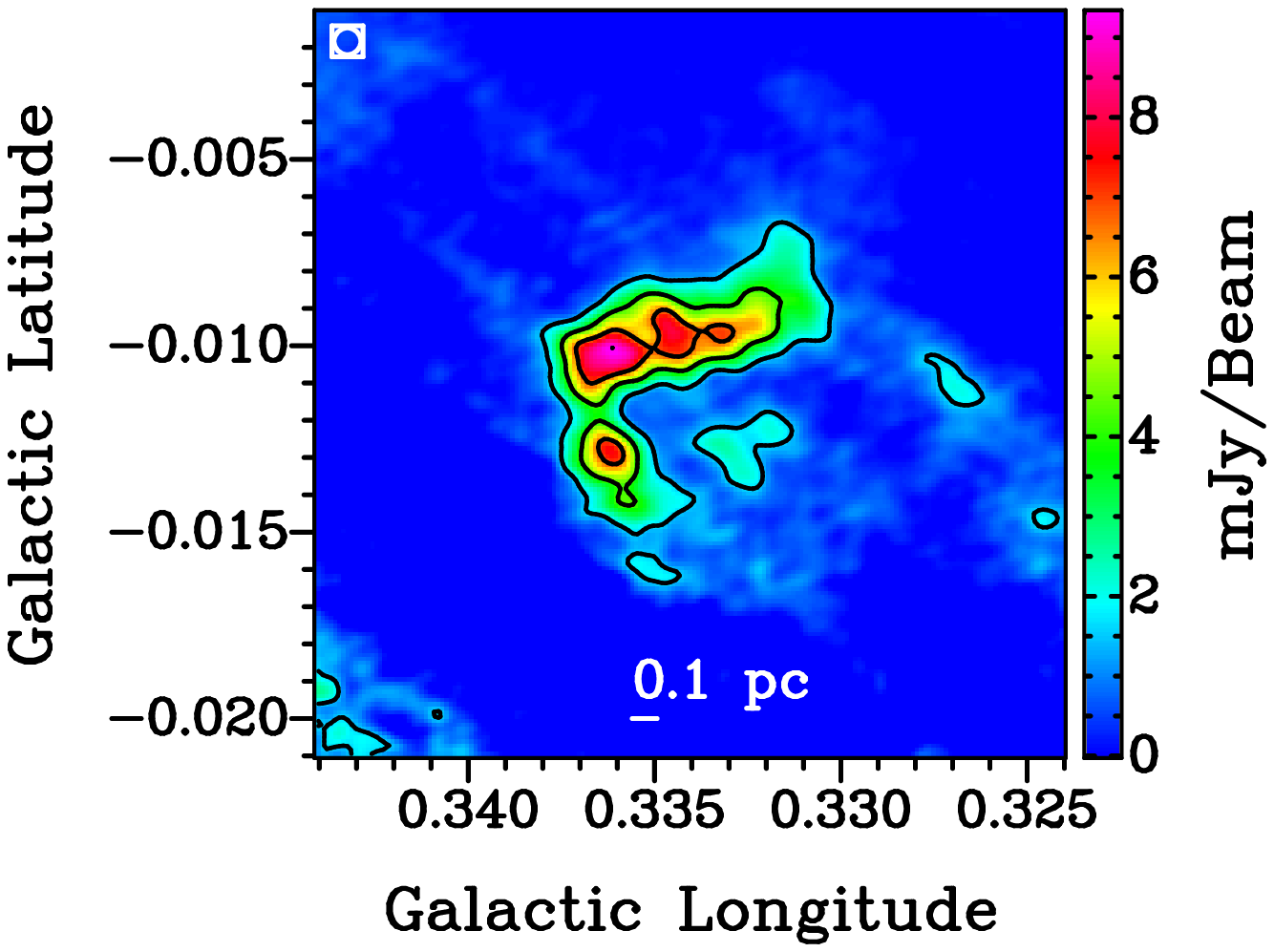}}		
      \subfloat[Source E]{\includegraphics[angle=270,width=6cm]{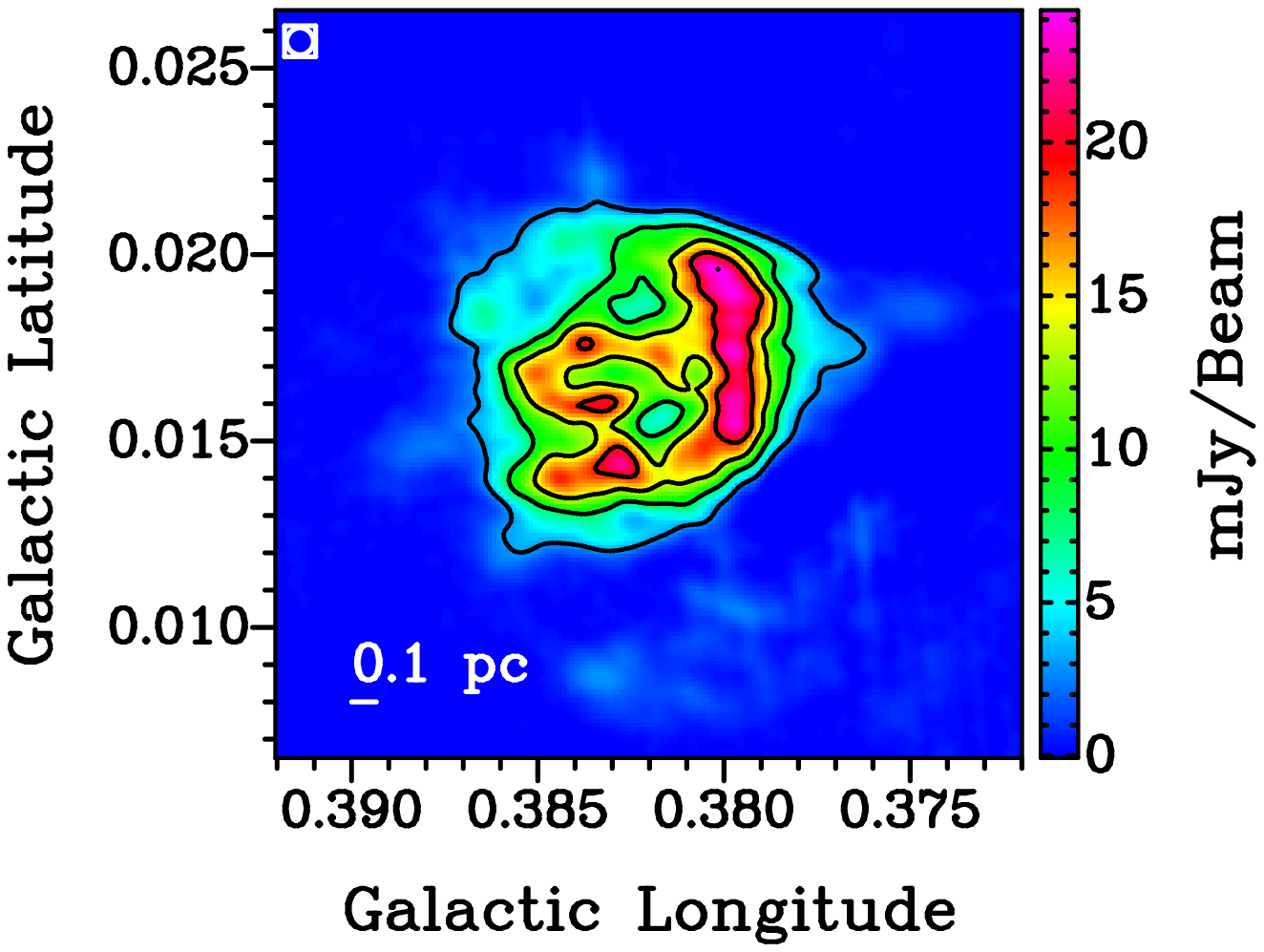}}
   \caption{a) 870 $\mu$m dust emission of the dust ridge (see Fig. \ref{DRSubmmTM}). The contours present the 1.4 GHz map of \citet{Yusef-Zadeh2004} with contour levels of 3\% to 10\% in steps of 1\%. The circles show the VLA primary beams at 8.4 GHz at the three observed positions. The sizes and positions of the yellow boxes correspond to the sizes and positions of the radio source images (b-f). b) - f) The images show the emission of the five detected radio sources at 8.4 GHz with contour levels of 20\% to 100\% in steps of 20\% (1\% corresponds to 1.4 $\sigma$, 0.2 $\sigma$, 0.3 $\sigma$, 0.6 $\sigma$, and 1.0 $\sigma$ in Fig. \ref{DR+RS}b, c, d, e, and f, respectively).}
   \label{DR+RS}
\end{figure*}

\begin{table*}
\begin{minipage}[t]{\textwidth}
\caption{Positions of the detected radio continuum sources (Col. 2, 3), the integrated flux densities at 8.4 GHz and 5 GHz (Col. 4,5), the spectral index (Col. 6), obtained from the integrated flux densities at 8.4 and 5 GHz, and the angular radii $\Theta_R$ (Col. 7). The uncertainty of the absolute flux calibration is less than 10\%, as determined from observations of the quasar 1730--130.}
\label{FluxParameterSources}
\centering
\renewcommand{\footnoterule}{} 
  \begin{tabular}{|c|c|c|c|c|c|c|c|}\hline
	Source 	& R.A. (J2000)  	& DEC (J2000)  &  S$_{\textnormal{8.4 GHz, Int.}}$ & S$_{\textnormal{5 GHz, Int.}}$\footnote{\citet{Becker1994}} & $\alpha$ (S$_\nu$ $\propto$ $\nu^\alpha$) & $\Theta_{R}$ & Comments    \\
			& [hh mm ss.ss] 	& [d $\arcmin\arcmin$ $\arcsec\arcsec.\arcsec\arcsec$] & [mJy] & [mJy] &  & [$\arcsec$] & \\ \hline
A & 17 46 07.35 & $-$28 45 31.21 & 180 $\pm$ 2 & 154 & 0.3 & 5.1 &  S$_{\textnormal{1.4 GHz, Peak}}$\footnote{\citet{Becker1994}} $\sim$ 90 mJy~beam$^{-1}$\\
B & 17 46 04.09 & $-$28 42 33.17  & 9 $\pm$ 1 & < 9 & > -0.2 & 1.8 &  \\
C & 17 46 06.15 & $-$28 39 46.72  & 10 $\pm$ 1& < 9 & > 0 & 1.8 & \\
D & 17 46 27.56 & $-$28 39 15.25 & 145 $\pm$ 20 & 134 & 0.2 & 14.1\footnote{at 5 GHz \citep{Becker1994}}  & S$_{\textnormal{1.4 GHz, Peak}}$ $\sim$ 60 mJy~beam$^{-1}$ \\
E & 17 46 28.93 & $-$28 36 14.80 & 886 $\pm$ 57 & 1417 & -0.9 & 32.0& S$_{\textnormal{1.4 GHz, Peak}}$ $\sim$ 51 mJy~beam$^{-1}$ \\\hline
\end{tabular}
\end{minipage}
\end{table*}

\begin{table*}
\begin{minipage}[t]{\textwidth}
\caption{Measured integrated flux densities at 8.4 GHz, S (Col. 2), and the angular radii $\Theta_R$ (Col. 3) as well as the calculated results for the electron densities n$_e$ (Col. 4), the emission measures EM (Col. 5) and the masses of the \ion{H}{ii} regions M$_{\ion{H}{ii}}$ (Col. 6). The last column contains the spectral types of the stars, assuming one single ZAMS star exciting the \ion{H}{ii} region. These values are calculated based on the assumption of a spherically symmetric HII region. Parameters of source D could not be determined, since the source is elongated.}
\label{PhysParameterSources}
\centering
\renewcommand{\footnoterule}{} 
  \begin{tabular}{|c|c|c|c|c|c|c|c|c|c|c|c|}\hline
	Source  & S$_{\textnormal{8.4 GHz, Int.}}$ & $\Theta_{R}$ & n$_e$ & EM & M$_{\textnormal{\ion{H}{ii}}}$    & spectral type    \\
			 & [mJy] & [$\arcsec$] & [cm$^{-3}$] & [10$^{5}$ cm$^{-6}$ pc] & [M$_{\sun}$] & (single ZAMS star\footnote{Obtained from \citet{Panagia1973}})     \\ \hline
A  & 180 & 5.1 & 2050 & 11.5 & 0.8 & O9-O9.5  \\
B  & 9 & 1.8 & 2200 & 4.7 & 0.04 & B0-B0.5  \\
C & 10 & 1.8 & 2330 & 5.3 & 0.04 & B0-B0.5  \\
D & 145   & & & & &\\
E  & 886  & 32.0& 290 & 1.4 & 28 & O6.5-O7 \\\hline
\end{tabular}
\end{minipage}
\end{table*}

\begin{figure*}[htbp]
     \centering
    \includegraphics[width=0.8\textwidth]{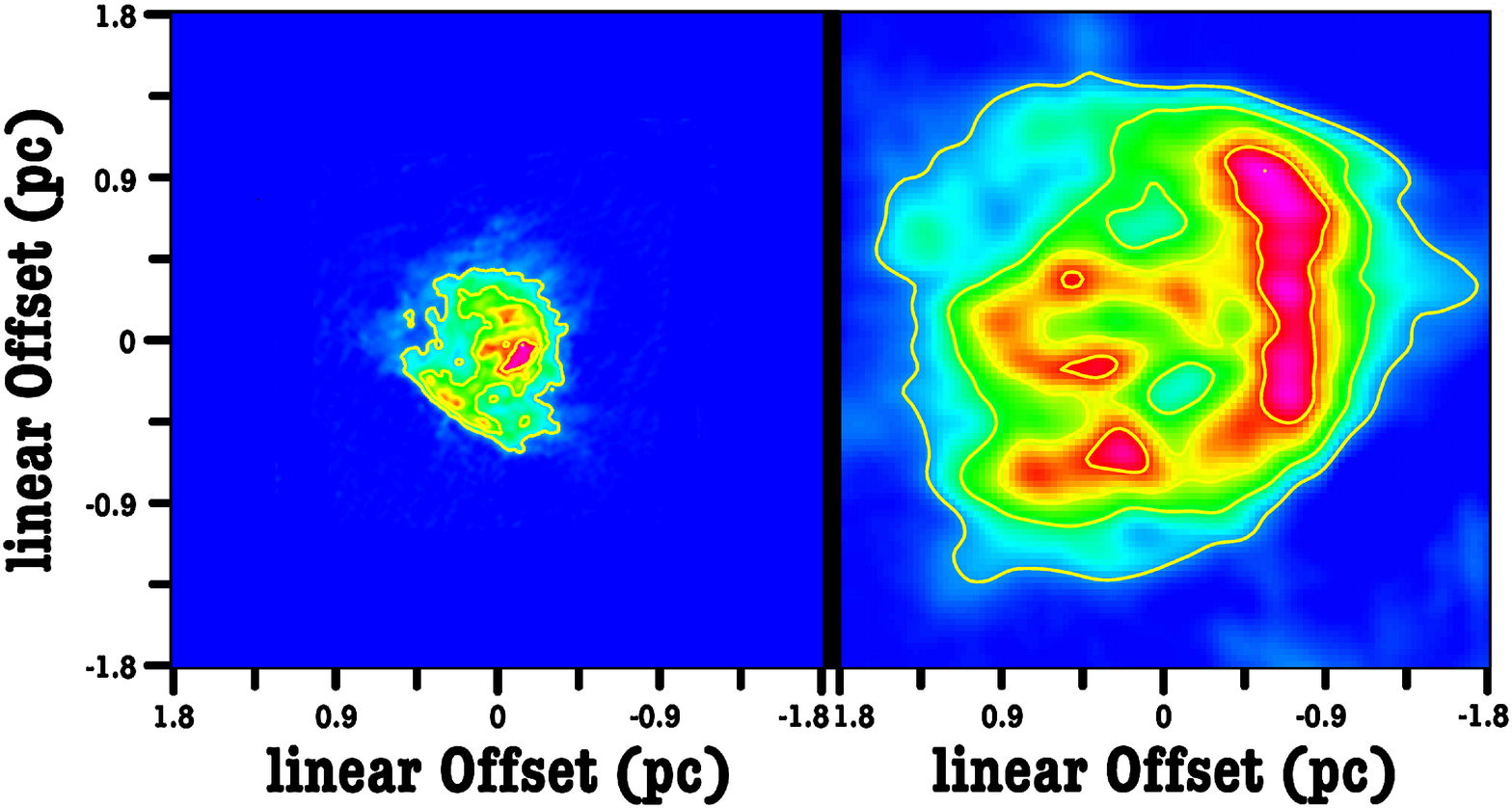}
   \caption{Left panel: The Orion Nebula (M42) at 8.4 GHz, taken with the VLA in D-array \citep{Shepherd2001}. Right: panel: Source E at 8.4 GHz. The contour levels are 20\% to 100\% in steps of 20\% in both panels (1\% corresponds to 1.4 $\sigma$ and 1.0 $\sigma$ in the left and the right panels, respectively).}
   \label{Orion-VLA}
\end{figure*}

Fig. \ref{DR+RS} shows that the primary beams of the VLA observations cover only the dust ridge clouds a -- d. Thus, we cannot make a statement about radio sources in clouds e and f from our observations. However, there are no 5 GHz or 1.4 GHz sources from the catalog of \citet{Becker1994} located in clouds e and f. In addition, radio emission is only detected towards the known \ion{H}{ii} region Sgr B1 in the south of clouds e and f in the 1.4 GHz map of \citet{Yusef-Zadeh2004} and in VLA archival data at 4.8 GHz. We, therefore, exclude the existence of \ion{H}{ii} regions in the clouds e and f.

A total of five radio sources are detected in the three final 8.4 GHz radio images. We fitted the sources in the primary beam corrected images with Gaussian models and determined their positions and angular sizes as well as their peak and integrated flux densities. We calculated the spectral indices of the radio sources from the integrated flux densities at 8.4 and at 5 GHz \citep[from the catalog of][]{Becker1994}. Since \citet{Becker1994} list only the 1.4 GHz peak flux densities of the counterparts in their catalog, integrated flux density values at 1.4 GHz could not be included in the determination of the spectral index.

The image centered on POS1 contains three sources (labeled A, B, and C in Fig. \ref{DR+RS}) which are all located offset from molecular cloud M0.25+0.012 to SWW, W and NE, respectively. Of these three sources only the southernmost source (A) is associated with 870~$\mu$m dust emission, as well as a water and a methanol maser (see Section \ref{MaserDR} and Fig. \ref{DRSubmmTM}). This source is also detected at 1.4 and 5 GHz \citep{Becker1994}. Its spectral index is $\alpha$ $\sim$ +0.3, supporting the identification of this source as an \ion{H}{ii} region.

The second southernmost source (B) was also detected by \citet{Lis1994b} who measured a peak flux density of 7.4 mJy~beam$^{-1}$ and an integrated flux density of 14.6 mJy at 8.4 GHz. They doubted that this source has an extragalactic nature but instead assumed that they detected a compact \ion{H}{ii} region due to its relatively high flux density. The Gaussian fitting of our data results in a peak flux density of (5.56 $\pm$ 0.58) mJy~beam$^{-1}$ and an integrated flux density of (9.27 $\pm$ 1.42) mJy. Assuming similar error values for the results of \citet{Lis1994b}, our results are consistent within 3 $\sigma$. 

The third detected source C is a small, nearly circularly shaped source which is also observable in the second field POS2. Its integrated flux density is even higher than the integrated flux density of source B. Source B and C are not detected in the 1.4 GHz map of \citet{Yusef-Zadeh2004} or in the 5 GHz catalog of \citet{Becker1994}. We determined lower limits for the spectral index of -0.2 (source B) and 0 (source C), using the 5 GHz detection limit of 9 mJy as upper limit. The spectral indices suggest an interpretation of the two radio sources as \ion{H}{ii} regions.

POS2 contains two more sources besides source C, which lie close to the edge of the image in the area of higher noise values. Source D has a peculiar shape, it is not circular or nearly circular like the other sources, but is strongly extended in the north-west to south-east direction. Because of the elongation of the source together with its position in the field, one could assume that this source is spurious, but since the 1.4 GHz map of \citet{Yusef-Zadeh2004} shows emission at this location and a 5 GHz continuum source \citep{Becker1994} coincides with this object, we conclude that this source is real. An explanation of the peculiar shape of this object could be the blending of several circular shaped sources. The spectral index of 0.2 supports the identification of this source as an \ion{H}{ii} region.

Source E is also observed in two fields (POS2, POS3) and has similar shapes in both fields. Of all the five sources this source exhibits the strongest emission with an integrated flux density of 0.89 Jy. This source was also detected in the 1.4 GHz map of \citet{Yusef-Zadeh2004} and as a 5 GHz continuum source by \citet{Becker1994} who identified it as an \ion{H}{ii} region. The spectral index of this source is -0.9 which is inconsistent with an \ion{H}{ii} region. However, the source is located in the higher-noise region at the edge of the field of view in both fields which could explain a loss of emission for this source. \citet{Anderson2011} determined an integrated flux density of 1513 $\pm$ 53 mJy at 8.7 GHz for this source which results in a spectral index of 0.1, consistent with an \ion{H}{ii} region.

Although five sources are found in the three fields, no radio continuum sources at 8.4 GHz could be detected within the dust ridge molecular clouds indicating that no high-mass star formation is taking place or that the high-mass star formation is in an early phase, where an \ion{H}{ii} region cannot yet be observed.

The positions of the five sources (indicated with an upper-case letter, Col. 1), the integrated flux density values at 8.4 and 5 GHz, the spectral index from the integrated flux densities at 8.5 and 5 GHz, and the angular radii $\Theta_R$ are listed in Table \ref{FluxParameterSources}.

Assuming that all these five sources have a Galactic origin and that they all are \ion{H}{ii} regions, excited by one zero-age main sequence star each, the spectral type of the stars that excite these regions can be estimated. This assumption is only questionable for source C since this source was not detected at other radio frequencies.

In order to determine the masses of the \ion{H}{ii} regions $M_{\ion{H}{ii}}$, we calculate the electron density $n_e$ and the emission measure $EM$ for the sources A, B, C, and E (see Appendix \ref{appendix}). Since the assumption of circular symmetric \ion{H}{ii} regions forms the basis of these calculations, they have not been conducted for source D. For the remaining sources, an electron temperature $T$ of 10$^4$ K and a distance of 8.5 kpc are assumed. The results of these calculations are listed in Table \ref{PhysParameterSources}. Col. 4 and 5 contain the electron densities and the emission measure values. The masses of the \ion{H}{ii} regions are shown in Col. 6. The spectral type of the stars is determined by comparing the number of Lyman continuum photons with the values of table II in the publication of \citet{Panagia1973} (shown in Col. 7). However, we cannot exclude that any of the detected sources are foreground objects which would mean that we overestimate their masses and the spectral types of the exciting stars.

The four sources show a wide range of \ion{H}{ii} masses from 0.04 to 28 M$_\odot$, probably being excited by zero-age main sequence (ZAMS) stars with spectral types earlier than B0.5. 

The most massive \ion{H}{ii} region (source E) needs for its excitation an O6.5 V to O7 V star. 
To put this source in perspective, let us compare it to the well studied archetypical compact \ion{H}{ii} region M42, the Great Nebula in Orion. The premier exciting source of M42 is $\Theta_1$~Ori~C, which has spectral type O7 V \citep{Stahl2008}, and is associated with the nearest site of active star formation \citep{Genzel1989}  at a distance of 414 pc \citep{Menten2007}. The left panel of Fig. \ref{Orion-VLA} shows an image of the Orion nebula at 8.4 GHz, taken with the VLA in compact D-array \citep{Shepherd2001}. The integrated flux density of M 42 at 8.4 GHz is 415 Jy \citep{Shepherd2001}.  The right panel shows source E at 8.4 GHz for comparison on the same physical scale. Source E is only slightly larger than the Orion nebula. Scaling the flux density of the Orion nebula to the same distance as source E yields a total flux density of 0.98 Jy, which is comparable to the integrated flux density of source E. Therefore, if we would move the Orion nebula to a Galactic Center distance of 8.5 kpc, we would see an \ion{H}{ii} region similar to source E. While the massive pre-main sequence stars in Orion are easily detectable due to the closeness to the Sun, the large distance to source E prevents the detection of the exciting stars. Thus, the only sign of the massive star formation in source E is the strong radio continuum emission. Needless to say, given the luminosities of its members and the pervasive high extinction, it would be completely impossible to find any trace of the $\sim 2$ million year old, $\sim 2000$ member Orion Nebula Cluster at optical wavelengths, had it been placed in the Galactic center region.

We can conclude that our data show evidence for massive star formation in a more evolved state taking place at the periphery of the dust ridge region, but outside its molecular clouds. The 8.4 GHz  3-$\sigma$ sensitivity limit of 0.15--0.25 mJy~beam$^{-1}$ in a $2\farcs{6}$ beam can be translated to a Lyman continuum luminosity of $\sim$ 2.4--5 $\cdot$ 10$^{45}$ s$^{-1}$ which corresponds to a single B0.5--B1 ZAMS star. Thus, we can exclude the existence of ZAMS star with spectral types earlier than B0.5 in the dust ridge clouds.

\subsection{Infrared Emission in the Dust Ridge}

\begin{figure*}
	\centering
   \includegraphics[angle=270,width=15cm]{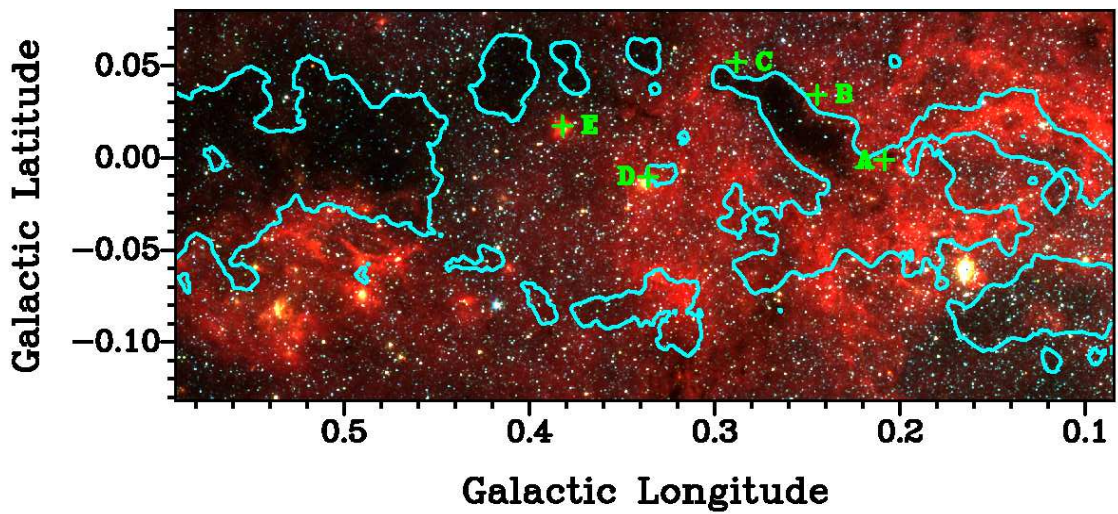}
		\caption{This picture shows the dust ridge in absorption against the infrared background (GLIMPSE RGB image of the dust ridge: blue = 3.6 $\mu$m, green = 4.5 $\mu$m, red = 8.0 $\mu$m). The contour shows dust emission at 20\% of the peak, with 1\% corresponding to 2 $\sigma$. It is clearly visible that a strong correlation exists between the distribution of the dust emission and the shape of the infrared dark clouds. The green crosses and the upper case letters mark the positions of the detected radio sources. At the positions of the radio sources A, D, and E strong infrared emission is detected.}
	\label{DustRidgeInfrared}
\end{figure*}

Observations of the dust ridge at infrared wavelengths give another possibility to find newly-born stars in the clouds since these stars heat the surrounding dust which then reemits this energy at infrared wavelengths.

Fig. \ref{DustRidgeInfrared} shows the dust ridge in a GLIMPSE false-color image. The contours display the distribution of the dust emission at 870 $\mu$m. The correspondence between the contours and the dark parts in the image in the northern part of the dust ridge is remarkable. The clouds in the southern part of the dust ridge are either not dense enough to absorb a large fraction of the diffuse infrared background or are located behind the material that emits the warm infrared background.

In addition, the infrared map shows emission at the positions of the three radio sources A, D, and E (marked with green crosses in Fig. \ref{DustRidgeInfrared}). Since no maser emission is detected in the last two sources, they might be in a more advanced stage in the star forming process.

\section{Conclusion}

Observation of the dust ridge at 870 $\mu$m reveal several dust clouds ordered along a narrow ridge between the radio continuum source G0.18-0.04 and Sgr B2. The temperature of the dust condensations is very low, between 15 and 22 K, which indicates that if high-mass star formation is taking place in the clouds, it is in a very early stage. Mid-infrared observations of the dust ridge from the GLIMPSE survey show the dust ridge clouds in absorption in front of the diffuse infrared background. The agreement between the infrared dark clouds and the dust emission is remarkable.

The mass of the dust clouds are very high, ranging from 13000 M$_{\sun}$ to over 150000 M$_{\sun}$, making this region one of the most massive reservoirs of molecular material near the Galactic Center. Due to the large masses and the low temperatures one can assume that a non-negligible part of the future star formation in the central molecular zone will take place in this region.

The detection of Class II methanol masers in two of the dust ridge clouds are the only sign that massive stars are born in the dust ridge. Except a weak water maser, no sign of ongoing star formation was found in the massive dust ridge cloud, M0.25+0.012.

Observation of the dust ridge region at 8.4 GHz resulted in the detection of five radio sources, which are probably all excited by young massive stars. However, the sources are found outside the massive dust clouds and only little dust can be observed at the position of these sources, indicating that the star formation at these locations is in a more evolved state than in the clouds. The presence of stars with spectral types earlier than B0.5 in the dust ridge clouds is excluded by our observations.

M0.25+0.012 and cloud e are by far the most massive clouds in the dust ridge. \citet{Longmore2012} identified M0.25+0.012 as a possible precursor of a young massive cluster like the Arches cluster due to its position in the radius-mass plot. The radius and mass of cloud e are comparable to the values of M0.25+0.012. Assuming that cloud e is gravitationally bound in the boundaries as defined in Section \ref{DustDR}, its position in the radius-mass plot would be very close to the position of M0.25+0.012 and this cloud could therefore also form a massive young cluster in the future. While M0.25+0.012 does not show any sign of star formation, cloud e contains a methanol maser which indicates ongoing star formation in this cloud. We will use recent APEX spectral line observations of the clouds a, d, and e in combination with data from the MALT90 survey to investigate the difference in chemistry between the clouds with and without signs of star formation and draw conclusions about the difference in the underlying physical conditions in the clouds.

\appendix
\section{}
\label{appendix}

Following \citet{Panagia1978}, the electron density n$_e$ of a circular symmetric \ion{H}{ii} region can be determined with the formula:
	\[n_e = 311.3 \cdot \left[\frac{S}{\textnormal{Jy}}\right]^{0.5} \left[\frac{T}{10^4~ \textnormal{K}}\right]^{0.25} \left[\frac{D}{\textnormal{kpc}}\right]^{-0.5} b(\nu,T)^{-0.5} \Theta_R^{-1.5} \textnormal{cm}^{-3}\]

\noindent with
 
\[b(\nu,T) = 1 + 0.3195 \cdot \log\left(\frac{T}{10^4~\textnormal{K}}\right) - 0.213 \cdot \log\left(\frac{\nu}{1~\textnormal{GHz}}\right).\]

\noindent where $S$ denotes the flux densities, $T$ the temperatures, and $\Theta_R$ the angular radii of the sources. $D$ denotes the source distance and $\nu$ the frequency of the observation. Furthermore, the emission measure EM can be estimated from the flux density $S$ and the angular radius $\Theta_R$ with the following formula, which can also be obtained from \citet{Panagia1978}:

\[EM = 5.638 \cdot 10^4 \left[\frac{S}{\textnormal{Jy}}\right] \left[\frac{T}{10^4~\textnormal{K}}\right] b(\nu, T) \Theta_R^{-2}~\textnormal{cm}^{-6}~\textnormal{pc}\]

In addition, \citet{TielensBuch} connects the emission measure with the number of Lyman continuum photons N$_{Lyc}$ that are necessary to produce this emission measure (a constant density nebula assumed):

\[EM = 1.6 \cdot 10^6 \left[\frac{n_e}{10^3~\textnormal{cm}^{-3}}\right]^{\frac{4}{3}} \left[\frac{N_{Lyc}}{5 \cdot 10^{49}~\textnormal{photons}~\textnormal{s}^{-1}}\right]^{\frac{1}{3}} \textnormal{cm}^{-6}~\textnormal{pc}\]

From the number of Lyman continuum photons the spectral type of the new-born star can be determined (assuming the existence of only one star in the \ion{H}{ii} region) and, together with the electron density, the mass M$_{\ion{H}{ii}}$ of the \ion{H}{ii} region can be calculated \citep{TielensBuch}:

\[M_{\ion{H}{ii}} \approx 80 \cdot \left[\frac{n_e}{10^3~\textnormal{cm}^{-3}}\right]^{-1} \left[\frac{N_{Lyc}}{5 \cdot 10^{49}~\textnormal{photons}~\textnormal{s}^{-1}}\right] \textnormal{M}_\odot\]
 
\bibliographystyle{aa}

\end{document}